\documentclass[english,aps,prd,twocolumn,superscriptaddress]{revtex4-1}

\usepackage{bm}
\usepackage{appendix}
\usepackage{graphicx}
\usepackage{amsmath}
\usepackage{amssymb}%
\usepackage{color}
\usepackage{lipsum}
\usepackage[colorlinks=true,citecolor=blue,urlcolor=blue]{hyperref}
\begin{document}

\title{Formation of a supergiant quantum vortex in a relativistic Bose-Einstein condensate driven by rotation and a parallel magnetic field}

\author{Tao Guo}
\affiliation{Department of Physics, Tsinghua University, Beijing 100084, China}

\author{Jianing Li}
\affiliation{Department of Physics, Tsinghua University, Beijing 100084, China}

\author{Chengfu Mu}
\affiliation{School of Science, Huzhou University, Zhejiang 313000, China}

\author{Lianyi He}
\affiliation{Department of Physics, Tsinghua University, Beijing 100084, China}

\date{\today}%

\begin{abstract}
Analysis based on the energy spectrum of noninteracting bosons shows that, under the circumstance of parallel rotation and magnetic field, charged bosons form a Bose-Einstein condensate because of the lift of the Landau level degeneracy by rotation [\textcolor{blue}{Y. Liu and I. Zahed, Phys. Rev. Lett. {\bf120}, 032001 (2018)}]. In this work, we study the interaction effect on the ground state of this Bose-Einstein condensate of charged bosons from the viewpoint of spontaneous symmetry breaking. We employ a minimal model for charged bosons with repulsive self-interaction. We find that the ground state of such a Bose-Einstein condensate is a supergiant quantum vortex, i.e., a quantized vortex with a large circulation. The size of the vortex is as large as the system size. The low-energy dispersion of the excitation spectra exhibits quadratic behavior, which is an anisotropic realization of the type-II Goldstone boson. Our study may give some implications to off-central relativistic heavy ion collisions, where large vorticity and magnetic fields can be generated. 
\end{abstract}

\maketitle

\section{Introduction}

Pions are the lightest hadrons of the strong interaction and are regarded as the pseudo-Goldstone bosons associated with the dynamical chiral symmetry breaking. As bosons, they may undergo Bose-Einstein condensation (BEC) in certain circumstances. The studies of quantum chromodynamics (QCD) at finite isospin chemical potential indicate that BEC of charged pions takes place when the isospin 
chemical potential exceeds the mass of charged pions \cite{PRL86:592,PRD66:034505,PRD97:054518}. It was proposed that BEC of pions may be formed  in compact stars~\cite{PRL29:382,PRL29:386,PRL30:1340,PRL31:257,PRD98:094510}, in heavy ion collisions~\cite{ZPA277:391,PRL43:1705,PLB316:226}, and in the early Universe~\cite{PRL121:201302,PRL126:012701,PRD104:054007}.

In relativistic heavy ion collisions,  large vorticity and magnetic fields can be generated. Theoretical studies predicted that noncentral collisions involve large angular momenta  in the range $10^3\sim10^5\hbar$~\cite{PRC77:024906,EPJC75:406,PRC93:064907,PRC94:044910}.  The global polarization of $\Lambda$ hyperon observed in off-central Au-Au collisions reported by the STAR Collaboration indicates a large vorticity with an angular velocity   $\Omega \approx (9\pm1)\times10^{21}$Hz $\sim0.05m_\pi$~\cite{NAT548:62}. Meanwhile, we expect that a large magnetic field $B$, parallel to the vorticity, is formed at the early stage of the collision. Numerical simulations indicated that the strength of the magnetic field reaches ${\rm e}B\sim m_\pi^2$~\cite{IJMPA24:5925,PRC85:044907,PPNP88:1}.  The state of QCD matter under the circumstance of parallel rotation and magnetic field (PRM) arises as an interesting theoretical issue. 
 
It was argued that PRM can induce BEC of charged pions or a pion superfluid phase based on the solution of the Klein-Gordon equation for \emph{noninteracting} pions in PRM~\cite{PRL120:032001}. The mechanism is simple but robust: The Landau level degeneracy of charged pions in a constant magnetic field is lifted by rotation, and the rotation then plays the role of a chemical potential. However, the true ground state of the pion superfluid with realistic pion-pion interaction is yet unknown. Under the circumstance of PRM, the ground state would be inhomogeneous, such as a quantized vortex or a vortex lattice.

In this work, we study the interaction effect on the ground state of such a pion superfluid. However, the realistic pion-pion interaction dictated by the chiral symmetry of QCD is rather complicated.  As a first step, we consider in the work a simple model of charged bosons,  a complex scalar field with repulsive self-interaction. We show that with repulsive self-interaction, the ground state of the BEC of charged bosons in PRM is a supergiant quantum vortex.  Quantum vortices are a type of topological defect exhibited in superfluids and superconductors. In a cylindrical vortex state, the macroscopic wave function of the condensate can be written as
\begin{equation}
\psi({\bf r})=f(\rho)e^{iw\theta},\ \ \ w\in \mathbb{Z},
\end{equation}
with the cylindrical coordinates ${\bf r}=(\rho,\theta,z)$. A stable vortex state in a quantum fluid is usually the singly quantized with winding number $w=1$ (e.g., superfluid at a given rotation rate). A vortex state with winding number $w\geq2$ is called a giant vortex since the radial profile behaves as $f(\rho)\sim \rho^{|w|}$ near the core. The giant vortex is usually unstable owing to its large energy cost 
($\sim w^2$). Instead, a lattice of singly quantized vortices will form. Searching for giant vortices is a longstanding topic in the research of quantum fluids~\cite{NAT404:471,PRL93:257002,PRL99:147003,PRL103:067007,PRL107:097202,PRA66:053606,PRL90:140402,PRA69:033608,PRA71:013605}. The BEC of charged bosons studied in this work provides an extreme example; the ground state is a supergiant quantum vortex with an extremely large winding number $w\gg1$.

This paper is organized as follows. In Sec. II we set up a minimal model for interacting charged bosons in PRM. In Sec. III we review the energy spectrum of noninteracting charged bosons in PRM, which shows that BEC of charged bosons can occur. 
In Sec. IV we study the BEC of interacting bosons in PRM from the viewpoint of spontaneous symmetry breaking and show that the ground state of such a BEC is a supergiant quantum vortex.  
The excitation spectra of the vortex state are studied in Sec. V. We summarize in Sec. VI.

\section{Minimal Model}

Recent lattice QCD calculations indicate that pions can still be treated as point particles for magnetic field strength ${\rm e}B\sim m_\pi^2$~\cite{PRD104:014504}. The realistic interaction between pions dictated by the chiral symmetry of QCD is rather complicated.  As a first step, we consider a simple model of charged bosons,  a relativistic complex scalar field with a repulsive self-interaction. The Lagrangian density is given by
\begin{equation}
\mathcal{L}=\left(\partial_\mu \Phi^*\right)\left(\partial^\mu\Phi\right) -m_\pi^2|\Phi|^2  - \lambda |\Phi|^4.
\end{equation}
For charged pions, the coupling constant may be set to be $\lambda=m_\pi^2/(2f_\pi^2)$ so that the $s$-wave pion-pion scattering length in the $I=2$ channel at the tree level, 
$a_{\pi\pi}=m_\pi/(16\pi f_\pi^2)$~\cite{PRL17:616}, is recovered. However, this is inadequate since the realistic interaction dictated by the chiral symmetry of QCD is complicated and the neutral pion
should be taken into account. In addition, in the presence of a magnetic field, new interaction will be induced. Our theoretical results in this work can only be applied
to a simple system of charged bosons.

We replace the system in a constant magnetic field along the $z$ direction, $\boldsymbol{B}=B\hat {z}$.  Furthermore, 
a global rigid rotation along the magnetic field is applied, with angular velocity $\boldsymbol{ \Omega}=\Omega\hat{z}$.  We consider the case ${\rm e}B>0$ and $\Omega>0$ without loss of generality. 
It is convenient to study the system in a rotating frame. The spacetime metric $g_{\mu\nu}$ of the rotating frame is given by
\begin{equation}
ds^2=(1-\Omega^2\rho^2)dt^2+2\Omega ydxdt-2\Omega x dydt-d{\bf r}^2,
\end{equation}
where ${\bf r}=(x,y,z)$ and $\rho=\sqrt{x^2+y^2}$. The cylindrical coordinates ${\bf r}=(\rho,\theta,z)$ will also be used in the following. The action of the system is given by
\begin{equation}
{\cal S}=\int d^4x\sqrt{-|g|} \left[g^{\mu\nu}\left(D_\mu \Phi\right)^*\left(D_\nu\Phi\right) -m_\pi^2|\Phi|^2  - \lambda |\Phi|^4\right],
\end{equation}
where $|g|=\det(g_{\mu\nu})$. The constant magnetic field enters the Lagrangian density through the covariant derivative $D_\mu=\partial_\mu +i {\rm e}A_\mu$, where $A_\mu$ is the vector potential in the rotating frame.  Since $\sqrt{-|g|}=1$, it is convenient to rewrite the action as ${\cal S}=\int d^4x{\cal L}$, with the Lagrangian density 
\begin{equation}\label{Lag-R}
\mathcal{L}=|(D_t+\Omega y D_x - \Omega x D_y)\Phi|^2 - |D_i\Phi|^2 -m_\pi^2|\Phi|^2  - \lambda |\Phi|^4.
\end{equation}

In the rest frame, it is convenient to use the symmetric gauge $A_\mu^{\rm R}=(0,By_{\rm R}/2,-Bx_{\rm R}/2,0)$ so that the rotational symmetry along the $z$-axis is manifested. The vector potential in the rotating frame is then given by $A_\mu=\left(-B\Omega \rho^2/2,By/2,-Bx/2,0\right)$ according to the coordinate transformation to the rotating frame $t_{\rm R}=t$, $\rho_{\rm R}=\rho$, $\theta_{\rm R}=\theta+\Omega t$. Note that an additional electric field ${\bf E} = B\Omega \mbox{\boldmath{$\rho$}}$ is induced in the rotating frame. However, according to the identity
$D_t+\Omega y D_x - \Omega x D_y = \partial_t + \Omega y \partial_x - \Omega x \partial_y$,
the induced electric field ${\bf E}$ cancels out automatically, indicating that the rotating frame corresponds only to a frame change with no new force~\cite{PRL120:032001}. Therefore, 
the Lagrangian density (\ref{Lag-R}) reduces to
\begin{eqnarray}
\mathcal{L} = |(\partial_t-i\Omega L_z)\Phi|^2 - |D_i\Phi|^2 -m_\pi^2|\Phi|^2  - \lambda |\Phi|^4,
\end{eqnarray}
where  $L_z \equiv -i(x\partial_y-y\partial_x) = -i \partial_\theta$ is the angular momentum along the $z$ direction.

\section{Free-Particle Picture}
In the absence of interaction ($\lambda=0$), the Klein-Gordon equation in PRM is given by
\begin{equation}\label{KGE1}
\left[-(\partial_t- i \Omega L_z)^2+K_{\rm 2D}+\partial_z^2 -m_\pi^2\right]\Phi(t,{\bf r})= 0,
\end{equation}
where the operator $K_{\rm 2D}$ is defined as
\begin{eqnarray}
K_{\rm 2D}=\frac{\partial^2}{\partial \rho^2}+\frac{1}{\rho}\frac{\partial}{\partial \rho} -\frac{L_z^2}{\rho^2}  - \frac{1}{4}{\rm e}^2B^2\rho^2 + {\rm e}BL_z.
\end{eqnarray}
Consider a cylindrical system with radius $R$.  The solution can be written as
\begin{eqnarray}
\Phi(t,{\bf r})=e^{-iE t + i p_z z +i l \theta}\varphi_{nl}(\rho),
\end{eqnarray}
where $p_z$ is the momentum along the $z$-direction and $l$ is the angular momentum quantum number. The solution of the radial part can be given by~\cite{PRL120:032001}
\begin{equation}\label{Ks}
\varphi_{nl}(\rho)  =  {\cal N}_{nl}\ \rho^{|l|} e^{-\frac{1}{4}{\rm e}B\rho^2}{_1F_1}\left(-a_{nl},|l|+1,\frac{{\rm e}B\rho^2}{2}\right),
\end{equation}
where ${\cal N}_{nl}$ is a normalization factor and ${_1F_1}$ is a confluent hypergeometrical function with the parameter
\begin{equation}
-a_{nl} = \frac{1}{2}(|l| - l +1)-\frac{1}{2{\rm e}B} [(E + \Omega l)^2 - p_z^2 - m_\pi^2].
\end{equation}
The energy levels $E=E_{nl}$ can be obtained by imposing a zero boundary condition at $\rho=R$, i.e.,
\begin{equation}
{_1F_1}\left(-a_{nl},|l|+1,\frac{{\rm e}BR^2}{2}\right)=0,
\end{equation}
where $a_{nl}$ is defined as the $(n+1)$th zero of left-hand side for a given $l$. 

For a large system with radius $R \rightarrow \infty$, $a_{nl}\rightarrow n$, and the function ${_1F_1}$ reduces to an associated Laguerre polynomial.
At $\Omega=0$, the energy spectrum recovers the Landau levels 
\begin{eqnarray}
E_n=\sqrt{p_z^2+m_\pi^2+{\rm e}B(2n+1)},
\end{eqnarray}
with $-n<l<N-n$. Here $N\equiv {\rm e}BR^2/2$ is the degeneracy of the Landau levels.
When a rotation is turned on, the degeneracy is lifted and the spectrum becomes $E_{nl}=E_n\pm\Omega l$. The term $\Omega l$ plays the role of a chemical potential. BEC takes place if the largest chemical potential $\Omega N$ 
exceeds the effective mass $\sqrt{m_\pi^2+{\rm e}B}$ in the lowest Landau level~\cite{PRL120:032001}. 

In the following, we will study the BEC of \emph{interacting} bosons from the viewpoint of spontaneous breaking of the U$(1)$ symmetry.

\section{Ground State with Interaction}

Now we turn on the interaction. It is useful to work in the imaginary-time formalism. The partition function of the system is given by
\begin{eqnarray}
\mathcal{Z} = \int [d\Phi^*] [d\Phi] \exp (-{\cal S}_{\rm E}), 
\end{eqnarray}
where the action reads
\begin{eqnarray}\label{action}
{\cal S}_{\rm E} =\int_X\left\{\Phi^*{\cal G}\Phi +\lambda |\Phi|^4\right\},
\end{eqnarray}
with  
\begin{eqnarray}
{\cal G}=m_\pi^2-(\partial_\tau -\Omega L_z)^2-K_{\rm 2D}-\partial_z^2.
\end{eqnarray}
Here $\int_X\equiv\int_0^\beta d\tau \int d^3 \mathbf{r}$, with $\tau$ being the imaginary time and $\beta$ being the inverse of the temperature.  We focus on the zero temperature limit ($\beta\rightarrow \infty$) in the following.
If the ground state is a Bose-Einstein condensate, the complex scalar field $\Phi (\tau, \mathbf{r})$ acquires a nonzero expectation value.
Therefore, we decompose the quantum field $\Phi (\tau, \mathbf{r})$  into its classical part $\psi(\mathbf{r})$ and its quantum fluctuation $\phi(\tau, \mathbf{r})$, i.e.,
\begin{equation}\label{field-decom}
\Phi (\tau,\mathbf{r}) = \psi(\mathbf{r}) +\phi (\tau,\mathbf{r}).
\end{equation}
While the condensate $\psi(\mathbf{r})$ is static, it can be inhomogeneous. The profile of the condensate is determined by minimizing the effective potential, which is rather hard to evaluate beyond the tree level. At the tree level, the treatment is in analogy to the Gross-Pitaevskii (GP) theory of nonrelativistic Bose-Einstein condensates~\cite{BOOK01,BOOK02}. The GP potential of the present system is given by
\begin{widetext}
\begin{equation}\label{gpl-fe}
U[\psi(\mathbf{r})]=\int d^3 \mathbf{r} \left[\psi^*({\bf r})\left(m_\pi^2-\Omega^2 L_z^2 -K_{\rm 2D}-\partial_z^2\right)\psi({\bf r}) +\lambda |\psi({\bf r})|^4\right]
\end{equation}
Furthermore, we assume that the condensate is homogeneous along the $z$ direction, $\psi({\bf r})=\psi(\rho,\theta)$. The GP potential per length along the $z$-direction reads
\begin{eqnarray}\label{gpl-fe1}
\mathcal{U}[\psi(\rho,\theta)]=\frac{U[\psi(\mathbf{r})]}{\int dz}=\int_{\rm 2D}\Big [\psi^* (\rho,\theta)\left(m_\pi^2-\Omega^2 L_z^2 -K_{\rm 2D}\right)\psi(\rho,\theta) + \lambda |\psi(\rho,\theta)|^4\Big],
\end{eqnarray}
\end{widetext}
where $\int_{\rm 2D}\equiv \int_0^{2\pi} d\theta \int_0^R \rho d\rho$. The minimization of ${\cal U}$ leads to a relativistic GP equation in PRM,
\begin{eqnarray}
\left[m_\pi^2-\Omega^2 L_z^2 -K_{\rm 2D}+2\lambda |\psi(\rho,\theta)|^2\right]\psi(\rho,\theta) =0.
\end{eqnarray}

The next task is to solve the 2D problem defined by $\mathcal{U}[\psi(\rho,\theta)]$. To perform a variational calculation, it is useful to expand 
$\psi(\rho,\theta)$ in terms of a complete set of basis functions. To this end, we note that the eigenfunction of the operator in the quadratic term, $m_\pi^2-\Omega^2 L_z^2 -K_{\rm 2D}$, is given by
\begin{equation}
F_{nl}(\rho,\theta)=\varphi_{nl}(\rho)\Theta_l (\theta),
\end{equation}
with eigenvalues 
\begin{equation}
K_{nl}=m_\pi^2-\Omega^2l^2+ {\rm e}B (2a_{nl} +|l|-l+1).
\end{equation}
Here $\Theta_l(\theta)=e^{il\theta}/\sqrt{2\pi}$ and $\varphi_{nl}(\rho)$ is the solution (\ref{Ks}) of the Klein-Gordon equation. Therefore, it is natural to expand $\psi(\rho,\theta)$ in terms of $F_{nl}(\rho,\theta)$, i.e.,
\begin{equation}\label{psi-e}
\psi(\rho,\theta)=\sum_{n=0}^{\infty}\sum_{l=-\infty}^{\infty}c_{nl}F_{nl}(\rho,\theta).
\end{equation}
The GP potential can be expressed in terms of the variational parameters $c_{nl}$ as
\begin{eqnarray}\label{U-cnl}
\mathcal{U}= \sum\limits_{nl} K_{nl} |c_{nl}|^2  + \lambda \int_{\rm 2D} \bigg|\sum_{nl}c_{nl}F_{nl}(\rho,\theta)\bigg|^4.
\end{eqnarray}

\begin{figure}[t]
\centering
    \vspace{-0.8cm}
    \includegraphics[width=0.36\textwidth,height=0.5\textwidth]{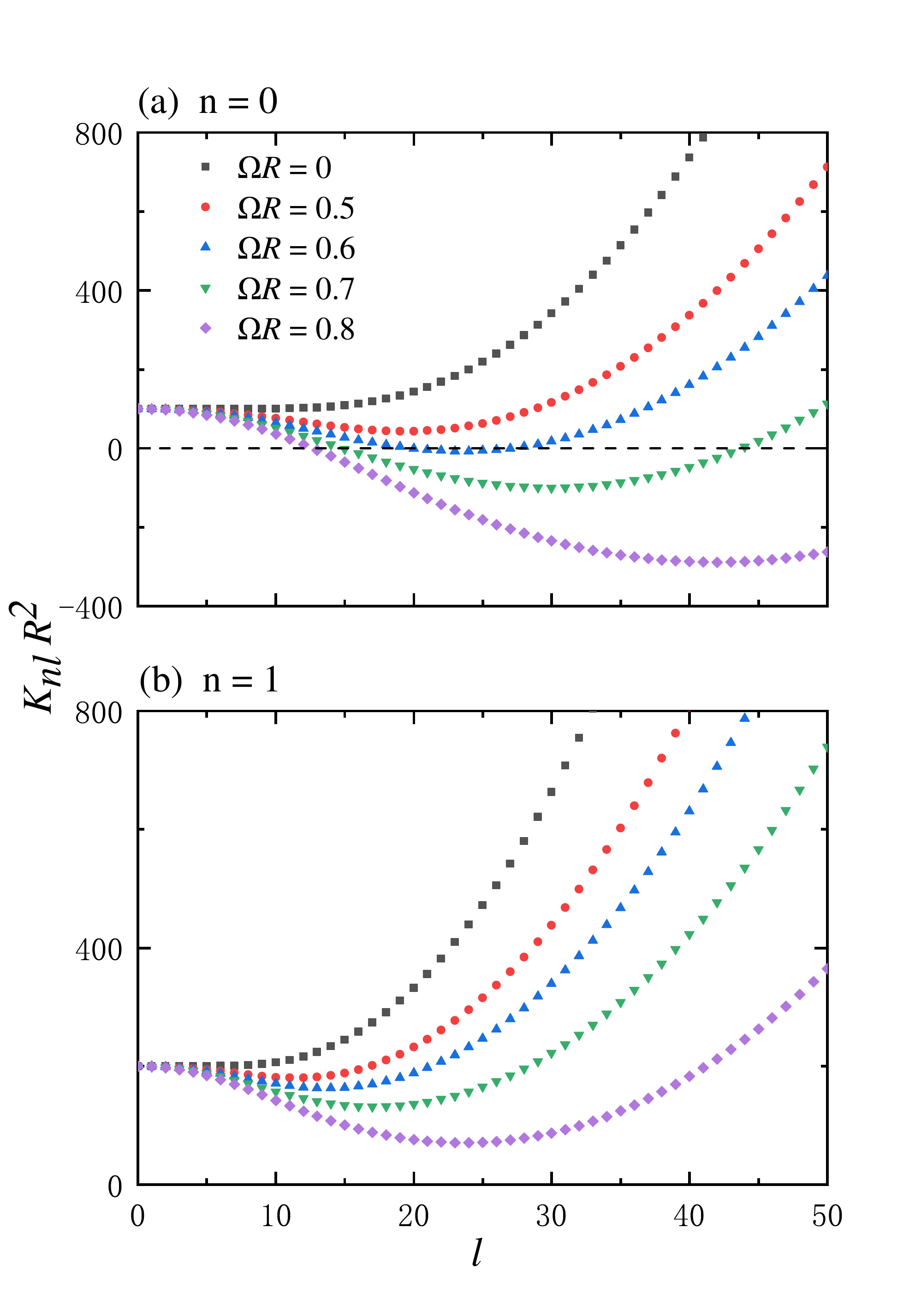}
    \vspace{-0.5cm}
\caption{\label{f1} $l$-dependence of the quantity $K_{nl}$ for different values of $\Omega$. Here we show the result for $n=0$ and $n=1$. The behavior for higher levels is similar.
In this plot we take ${\rm e}B=m_\pi^2$ and $N=25$. }
\end{figure}

\begin{figure}[t]
\centering
\begin{minipage}{9.2cm}
    \includegraphics[width=0.8\textwidth,height=0.99\textwidth]{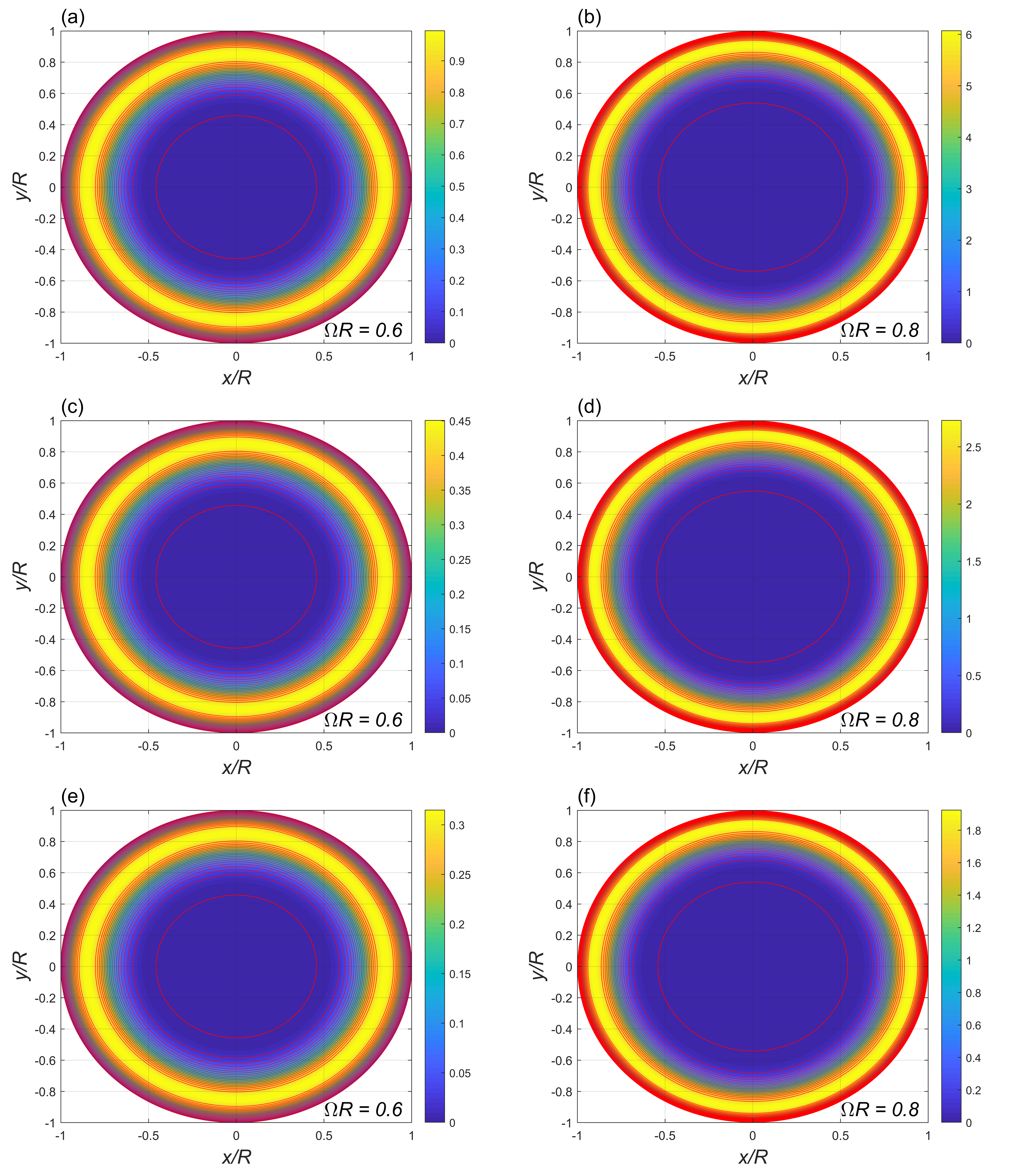}
\end{minipage}
\hspace{-0.5cm}

\caption{\label{f2} Profile of the condensate $|\psi(\rho,\theta)|$ in the $x-y$ plane for various values of the interaction strength: (a)(b)$\lambda=0.1$, (c)(d)$\lambda=0.5$, (e)(f)$\lambda=1$.
In this calculation we take ${\rm e}B=m_\pi^2$ and $N=25$. The color bars are in units of $m_\pi$.}
\end{figure}

\begin{figure}[t]
\centering
\vspace{-0.5cm}
    \includegraphics[width=0.52\textwidth,height=0.42\textwidth]{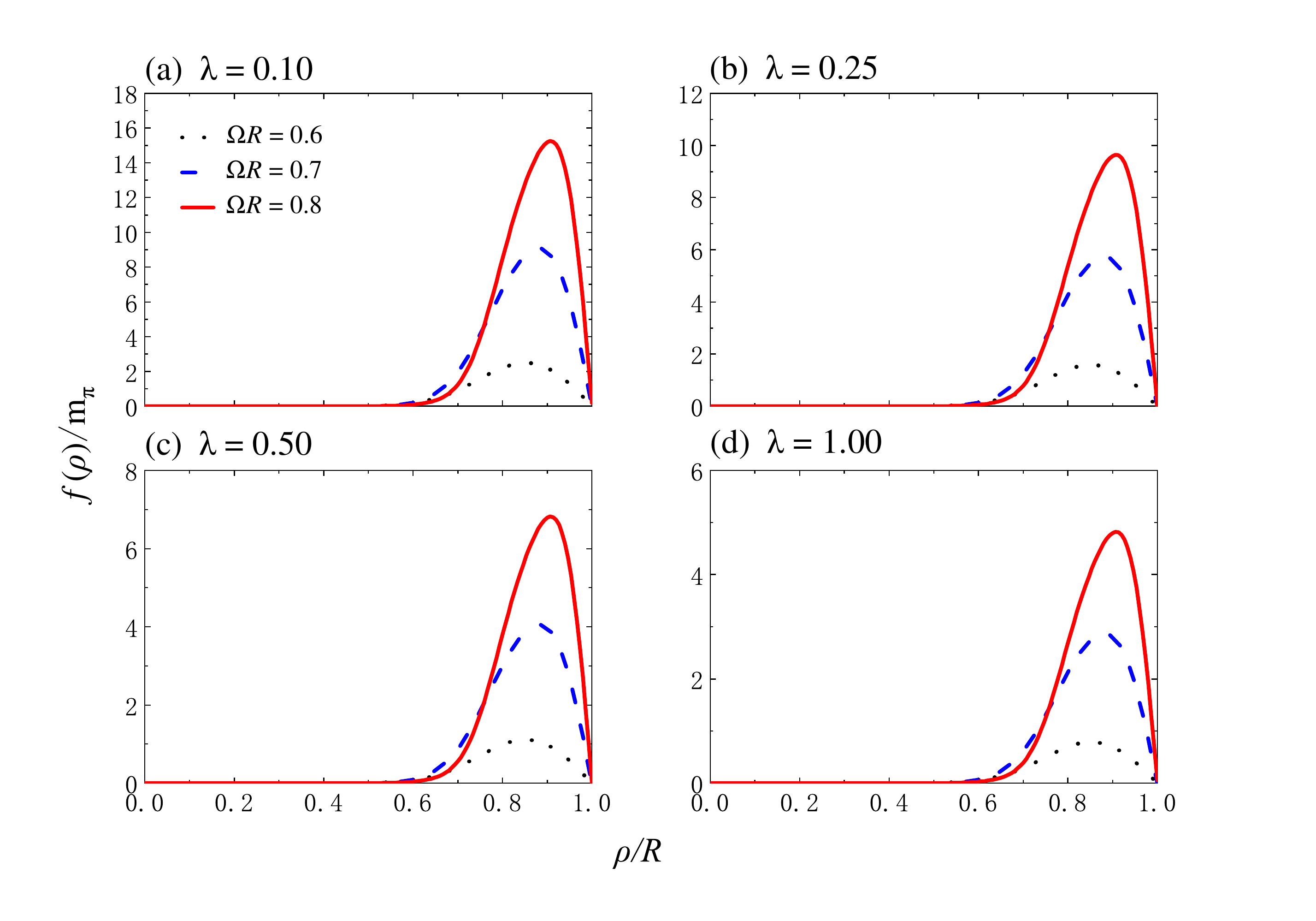}
\vspace{-1cm}

\caption{ \label{f3} Radial profile $f(\rho)$ of the condensate wave function for various values of interaction strength and rotation rate.
In this calculation we take ${\rm e}B=m_\pi^2$ and $N=25$.}
\end{figure}

The expression (\ref{U-cnl}) has a transparent quadratic plus quartic structure. Nonzero value of $\psi$ develops when at least one of the eigenvalues $K_{nl}$ becomes negative. Therefore, the critical 
condition for BEC is given by
\begin{equation}\label{BEC-CE}
\min_{nl}K_{nl}=0.
\end{equation}
Noting that $a_{nl}$ is always positive, we see clearly that the BEC is driven by the rotation: The quantity $\Omega l$ plays the role of an $l$-dependent chemical potential.  Because $a_{nl}$ is an increasing function of $n$, the critical condition for BEC will be first fulfilled for $n=0$. For negative $l$, $K_{nl}$ is an increasing function of $|l|$ and is hence always positive. Therefore, BEC occurs only for positive $l$. For positive $l$, we have $K_{nl}=m_\pi^2+{\rm e}B (2a_{nl}+1)-\Omega^2l^2$.  In Fig. \ref{f1}, we demonstrate the $l$-dependence of $K_{nl}$ at $l>0$. At large $l$, $a_{nl}$ goes faster than $l^2$. The competition between the decreasing term $-\Omega^2l^2$ and the increasing term $2{\rm e}Ba_{nl}$ results in a \emph{unique global minimum} at a certain quantum number $l$. For sufficiently large $\Omega$, this minimum for $n=0$ becomes negative and the BEC is induced. The critical angular velocity is given by
\begin{equation}
\Omega_c=\frac{\sqrt{m_\pi^2+{\rm e}B (2a_{0l_*}+1)}}{l_*}.
\end{equation}
Here $l_*$ is the location of the minimum that is exactly zero. Taking ${\rm e}B=m_\pi^2$, for $N=25$ and $N=100$, we have $l_*=20$ and $l_*=84$, respectively. As $N$ goes to infinity ($R\rightarrow\infty$), $l_*\rightarrow N$. The critical angular velocity approaches $\Omega_cR\simeq 2/\sqrt{N}$, which is vanishingly small. Thus the mechanism of rotation induced BEC in a magnetic field is quite robust. Note that the critical condition (\ref{BEC-CE}) is consistent with the analysis in the free-particle picture~\cite{PRL120:032001}.

\begin{figure}[t]

\centering
 \vspace{-0.5cm}
  \includegraphics[width=0.36\textwidth,height=0.5\textwidth]{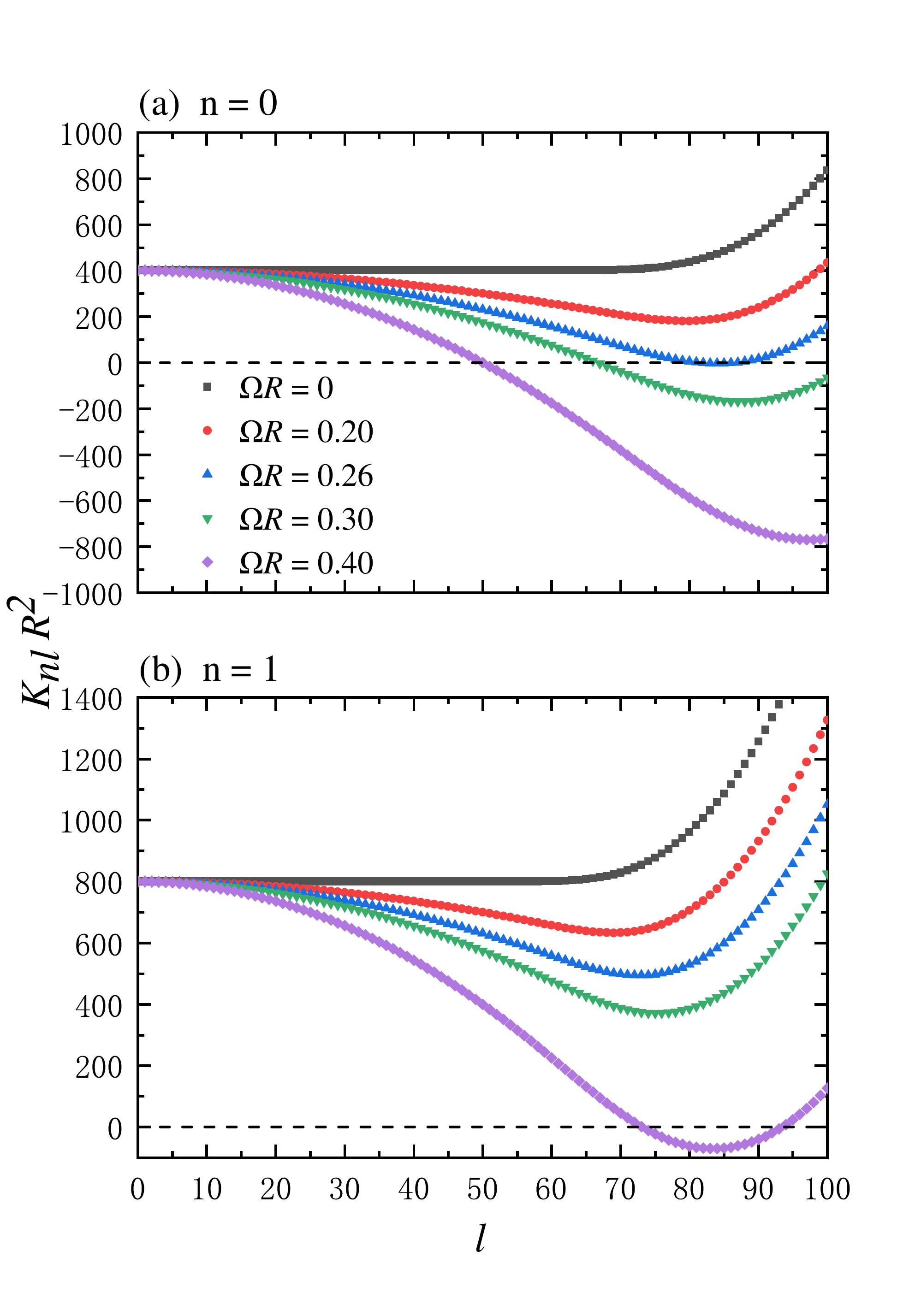}
 \vspace{-0.5cm}
\caption{\label{f4} $l$-dependence of the quantity $K_{nl}$ for different values of $\Omega$. Here we show the result for $n=0$ and $n=1$. 
In this plot we take ${\rm e}B=m_\pi^2$ and $N=100$. }
\end{figure}

\begin{figure}[t]
\centering
\begin{minipage}{9.2cm}
    \includegraphics[width=0.85\textwidth,height=1.05\textwidth]{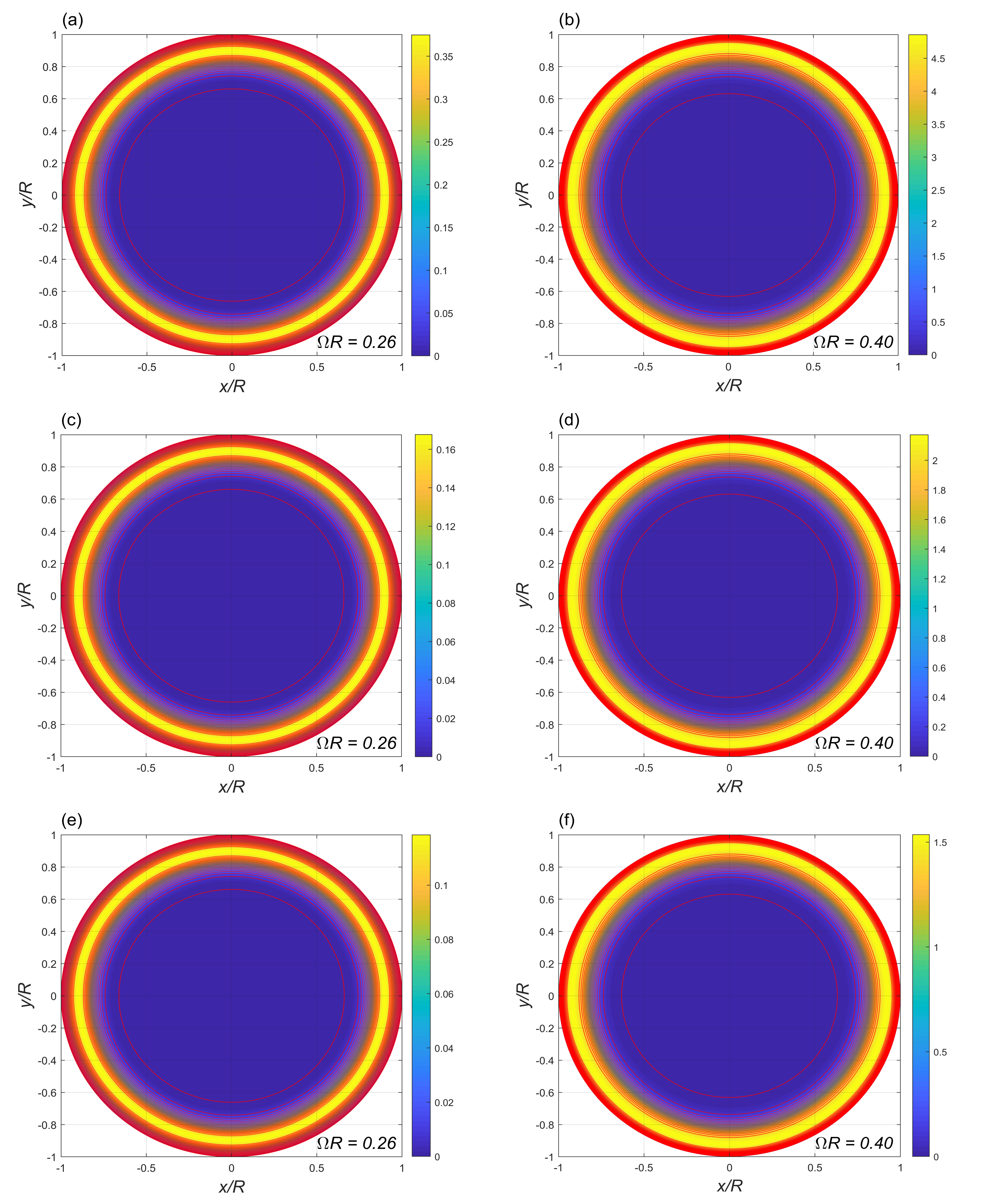}
\end{minipage}
\hspace{-0.5cm}

\caption{\label{f5} Profile of the condensate $|\psi(\rho,\theta)|$ in the $x-y$ plane for various values of the interaction strength: (a)(b)$\lambda=0.1$, (c)(d)$\lambda=0.5$, (e)(f)$\lambda=1$.
In this calculation we take ${\rm e}B=m_\pi^2$ and $N=100$. The color bars are in units of $m_\pi$.}
\end{figure}

\begin{figure}[t]
\centering
\vspace{-0.5cm}
\includegraphics[width=0.5\textwidth,height=0.4\textwidth]{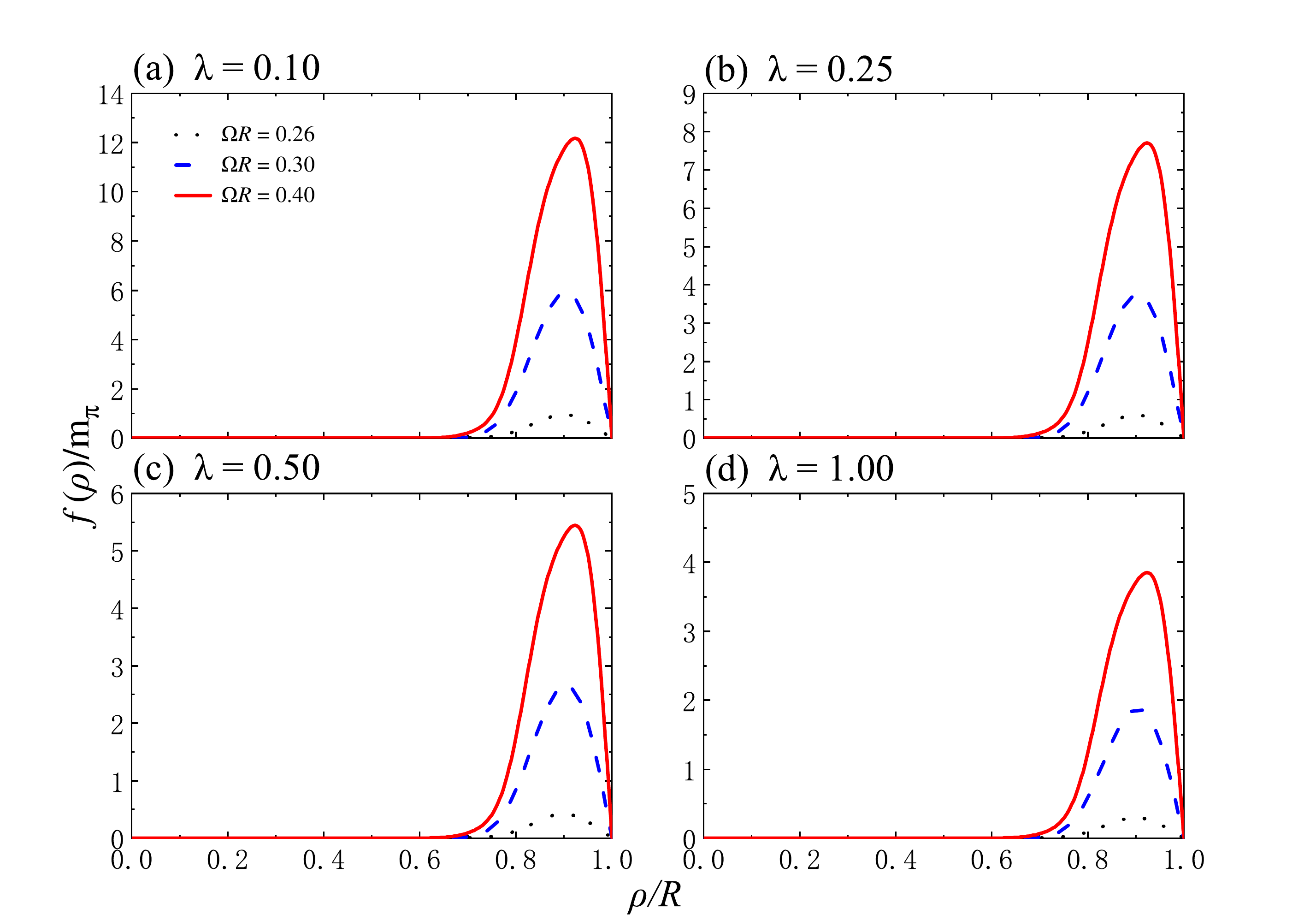}
\vspace{-0.5cm}

\caption{\label{f6} 
Radial profile $f(\rho)$ of the condensate wave function for various values of interaction strength and rotation rate. In this plot we take ${\rm e}B=m_\pi^2$ and $N=100$.}
\end{figure}

We then perform full variational calculations to determine the ground state in the BEC phase $\Omega>\Omega_c$. A typical numerical calculation has been performed for 
${\rm e}B=m_\pi^2$ and $N=25$, corresponding to a system size $R\simeq 10$fm. In this case, $\Omega_cR\simeq 0.59$. The two-dimensional profile of the condensate is shown in Fig. \ref{f2}. We find that it is always isotropic, indicating that the condensate wave function $\psi(\rho,\theta)$ is actually composed of a single $l$-component. This is supported by the results for the variational parameters:  
$c_{nl}$ is finite only for a certain quantum number $l=w$, while it is vanishingly small for all other values of $l$. Therefore, the condensate wave function can actually be expressed as 
\begin{equation}
\psi(\rho,\theta)=f(\rho) e^{iw\theta},
\end{equation}
with $f(\rho)=\sum_{n}c_{nw}\varphi_{nw}(\rho)$, corresponding to a vortex state.
This can be understood from the $l$-dependence of the quantity $K_{0l}$.  For $\Omega>\Omega_c$, the location of the \emph{unique minimum}, $l=l_0$,  moves to larger values ($l_0>l_*$) and the minimum becomes negative. As a result, if the interaction is sufficiently weak, the superposition (\ref{psi-e}) favors a single $l$-component with $w\simeq l_0$. Numerical results confirm this understanding. For example, the locations of the minima are $l_0=24,30,42$ for $\Omega R=0.6,0.7,0.8$, respectively. For $\lambda=1$, the winding numbers are respectively $w=23,29,39$. The small discrepancy comes from the combined effects of the interaction and the higher energy levels ($n\geq1$).

Therefore, in the presence of interaction, the ground state of the BEC is a supergiant vortex with winding number $w\gg1$. The condensate wave function takes the form (1).
The radial profile $f(\rho)$ is shown in Fig. \ref{f3}. Because of the large winding number, the size of the vortex is always as large as the system size.

We have also performed calculations for $N=100$ with the same magnetic field ${\rm e}B=m_\pi^2$, indicating a larger system size, $R\simeq 20$fm. We find the above conclusion remains. In this case, $\Omega_cR\simeq0.25$. In Fig. \ref{f4}, we show the  $l$ dependence of the quantity $K_{nl}$ for different values of $\Omega$. The qualitative behavior for the lowest level $n=0$ is the same as the case $N=25$, leading to similar results for the condensate profile (Fig. \ref{f5}). The radial profile of the condensate is shown in Fig. \ref{f6}.  We see that the size of the vortex is still as large as the system size.   The locations of the minima of $K_{0l}$ are $l_0=84,88,98$ for $\Omega R=0.26,0.3,0.4$, respectively. For $\lambda=1$, the winding numbers are, respectively, $w=84,86,93$, which are close to the values of $l_0$, as we expect.

\section{Excitation Spectra}

The elementary excitations in the BEC can also be calculated. They are quanta of the quantum fluctuation $\phi (\tau, \mathbf{r})$. Taking the quantum fluctuation into account, the action becomes ${\cal{S}}_{\rm E}=\beta U[\psi(\mathbf{r})]+ {\cal{S}}_{\rm fl}(\phi^\ast,\phi)$,
where the fluctuation contribution ${\cal{S}}_{\rm fl}$ can be obtained by substituting (\ref{field-decom}) into (\ref{action}). To calculate the excitation spectra, only the quadratic terms are relevant, which can be evaluated to be
\begin{equation}\label{flu1}
{\cal{S}}_{2}(\phi^\ast,\phi) = \int_X \left[\phi^*\left({\cal G} + 4\lambda |\psi|^2\right)\phi + \lambda \left(\psi^{* 2} \phi^2 + {\psi}^2 \phi^{*2}\right)\right].
\end{equation}
Considering the fact that the condensate takes the form $\psi({\bf r})=f(\rho)e^{iw\theta}$, we expand the fluctuation as
\begin{equation}
\phi (\tau,\mathbf{r}) = \sum_{n,Q} \tilde{\phi}_{n}(Q) e^{-i\omega_\nu \tau+ip_zz}F_{n,l+w}(\rho,\theta),
\end{equation}
where $Q=(i\omega_\nu,p_z,l)$ and $\omega_\nu = 2\pi \nu/\beta ~(\nu \in \mathbb{Z})$ is the boson Matsubara frequency.
Then we can write
\begin{equation}\label{flu3}
{\cal{S}}_{2} =  \frac{1}{2}\sum_{Q}\sum_{nn^\prime} \Lambda_{n}^\dagger(Q) {\bf M}_{n n^\prime}(Q) \Lambda_{n^\prime}(Q),
\end{equation}
where $\Lambda_{n}(Q)=(\tilde{\phi}_{n}(Q),\ \tilde{\phi}^*_{n}(-Q) )^{\rm T}$. The matrix ${\bf M}(Q)$ can be expressed as 
\begin{widetext}
\begin{equation}
{\bf M}_{nn^\prime}(Q)=  \left(
\begin{array}{cc}
{\cal K}_{n}(Q)\delta_{nn^\prime}+  {\cal A}_{nn^\prime}(l)  &  {\cal B}_{nn^\prime}(l)  \\[0.5em]
   {\cal B}_{nn^\prime}(-l)      &  {\cal K}_{n}(-Q)\delta_{nn^\prime}+  {\cal A}_{nn^\prime}(-l)
\end{array}\right),
\end{equation}
where $Q=(i\omega_\nu,p_z,l)$. The elements are given by
\begin{eqnarray}
&&{\cal K}_{n}(Q) = -[i\omega_\nu + (l+w)\Omega]^2 + p_z^2 + m_\pi^2 + \mathrm{e}B \left[2a_{n,l+w} +|l+w|-(l+w)+ 1\right],\nonumber\\
&&{\cal A}_{nn^\prime}(l) = \frac{2\lambda}{\pi} \int_0^R \rho d\rho \ f^2(\rho)\varphi_{n,l+w}(\rho) \varphi_{n^\prime,l+w}(\rho) ,\nonumber\\
&&{\cal B}_{nn^\prime}(l) = \frac{\lambda}{\pi} \int_0^R \rho d\rho \ f^2(\rho)\varphi_{n,l+w}(\rho) \varphi_{n^\prime,w-l}(\rho).
\end{eqnarray}
\end{widetext}
 The excitation spectra can be determined by $\det{\bf M}=0$ with the analytical continuation $i\omega_\nu\rightarrow E+i\epsilon$.

\begin{figure}[t]
\centering
\vspace{-0.5cm}
    \includegraphics[width=0.52\textwidth,height=0.21\textwidth]{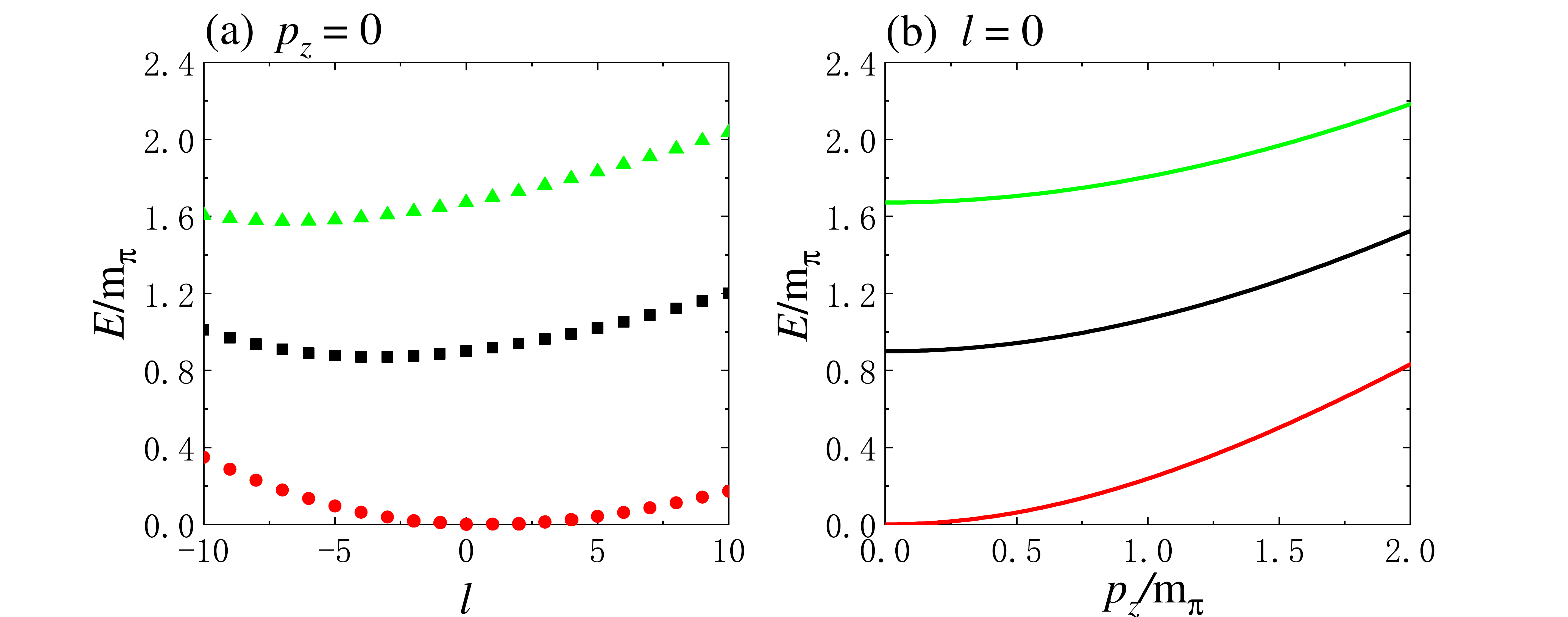}
\vspace{-0.5cm}
\caption{\label{f7} Energy of the elementary excitations as a function of $l$ for $p_z=0$ (a) and as a function of $p_z$ for $l=0$ (b) at $\Omega R=0.6$ for $\lambda=1$. The lowest three levels are shown in this plot. In this calculation we take ${\rm e}B=m_\pi^2$ and $N=25$.}
\end{figure}

Typical behavior of the excitation spectra is shown in Fig. \ref{f7} for $\Omega R=0.6$ and $\lambda=1$ with system size $N=25$. The lowest level of the spectra fulfills the Goldstone's theorem. Interestingly, the Goldstone mode in this system is an anisotropic generalization of the type-II Goldstone boson~\cite{PLB522:67,PRL88:111601,SYM2:609,PRL108:251602,PRL110:091601,PRX4:031057}; in the low-energy limit, the excitation energy $E$ shows a quadratic rather than linear dispersion.  Especially, for $l=0$, the dispersion relation of the lowest excitation can be well approximated as $E(p_z)\simeq \sqrt{p_z^2+(w\Omega)^2}-w\Omega$, where $w\Omega$ plays the role of a chemical potential. Thus the excitation energy exhibits a quadratic dispersion $E(p_z)\sim p_z^2$ for $p_z\rightarrow0$. 
This quadratic behavior is similar to the previously discussed type-II Goldstone boson at finite density~\cite{SYM2:609}. On the other hand, if the dynamical electromagnetic field is taken into account, the system becomes a superconductor and it is interesting to study how this gapless mode will be modified.

\section{Summary}

In summary, we have shown that under the circumstance of parallel magnetic field and rotation, the charged bosons get Bose condensed and the ground state is a supergiant quantum vortex with a 
winding number $w\gg1$. The condensate is almost located at the edge of the system, indicating that the size of the vortex is as large as the system size. The formation of supergiant vortex may give some implications to off-central heavy ion collisions. However, the realistic pion-pion interaction dictated by the chiral symmetry of QCD is rather complicated. In the future, we need to consider realistic 
pion-pion interactions and the chiral symmetry of QCD. It is also interesting to investigate the multi-pion Bose-Einstein correlations~\cite{NPA714:124,PRL93:152302,PLB696:328,PRC93:054908} and see how the Bose-Einstein condensation and the supergiant vortex (if still exist with realistic interaction) influence these correlations. It is also interesting to explore the relation between the hyperon polarization and the quantized vortex in such a pion superfluid~\cite{PRD96:096023}.

Finally,  the mechanism of forming super giant quantum vortices studied in this work is robust and general.  Since (pseudo) relativistic particles can be engineered in condensed matter and cold atom systems~\cite{NATP13:751,PRA102:033321,SCI372:271}, relativistic bosons under parallel magnetic field and rotation may also be engineered to explore the super giant quantum vortices.

\begin{acknowledgments}
We thank Lingxiao Wang, Ke-Ji Chen and Fan Wu for useful discussions.
This work was supported by the National Key R\&D Program (Grant No. 2018YFA0306503) and the National Natural Science Foundation of China (Grant No. 11890712).

\end{acknowledgments}

\bibliographystyle{apsrev4-1}

\end{document}